\documentclass[12pt]{iopart}

\newcommand{\prc}{Phys. Rev. C}
\newcommand{\prd}{Phys. Rev. D}
\newcommand{\prl}{Phys. Rev. Lett.}
\newcommand{\apj}{ApJ}
\newcommand{\apjl}{ApJL}
\newcommand{\aap}{A\&A}
\newcommand{\araa}{Ann. Rev. Astron. \& Astrophy.}

\usepackage{graphicx}
\usepackage{amssymb}
\usepackage[usenames]{color} 

\begin{document}

\title[Nucleosynthesis in neutrino-driven winds] {Nucleosynthesis of
  elements between Sr and Ag in neutron- and proton-rich
  neutrino-driven winds}

\author{A.~Arcones} 

\address{Institut f\"ur Kernphysik, Technische Universit\"at
  Darmstadt, Schlossgartenstra{\ss}e 2, D-64289 Darmstadt, Germany}

\address{GSI Helmholtzzentrum f\"ur Schwerionenforschung GmbH,
  Planckstr. 1 D-64291 Darmstadt, Germany}

\ead{almudena.arcones@physik.tu-darmstadt.de}

\author{J. Bliss}
\address{Institut f\"ur Kernphysik, Technische Universit\"at
 Darmstadt, Schlossgartenstra{\ss}e 2, D-64289 Darmstadt, Germany}

\begin{abstract}
Neutrino-driven winds that follow core collapse supernovae were thought
to be the site where half of the heavy elements are produced by the
r-process. Although recent hydrodynamic simulations show that the
conditions in the wind are not enough for the r-process, lighter heavy
elements like Sr, Y, and Zr can be produced. However, it is still not
clear whether the conditions in the wind are slightly neutron rich or
proton rich. Here we investigate the nucleosynthesis in
neutrino-driven winds for both these conditions and systematically
explore the impact of wind parameters on abundances. Our results
show the difficulty of obtaining a robust abundance pattern in
neutron-rich winds where also an over production of Sr, Y, and Zr is
likely. In proton-rich conditions, the abundances smoothly change when
varying wind parameters. Constraints for wind parameters and neutrino
energies and luminosities will soon become available by combining
nucleosynthesis studies, like the one presented here, with new and
future experimental data and observations.
\end{abstract}

\submitto{\JPG}
\maketitle


\section{Introduction}
\label{sec:intro}
Which elements heavier than iron are synthesized in neutrino-driven
winds?  Half of the nuclei heavier than iron are produced by the
r-process (rapid neutron capture process). This is characterized by
very fast neutron captures compared to beta decays and thus by the
production of extreme neutron-rich nuclei that eventually decay to
stability (for a review see \cite{arnould.goriely.takahashi:2007}).
The neutrino-driven winds that develop after successful supernova
explosions \cite{duncan.shapiro.wasserman:1986} were suggested as the
site of the r-process \cite{Meyer.Mathews.ea:1992,
  Woosley.etal:1994}. The very neutron-rich and explosive conditions
required by this process point to core-collapse supernovae, neutron
star mergers \cite{Lattimer77, Freiburghaus.Rosswog.Thielemann:1999},
and accretion disks \cite{Surman.Mclaughlin:2004}. In addition, the
number of these events times the amount of r-process material produced
by them has to account for the r-process isotopes in the solar system
and in the oldest observed stars \cite{Sneden.etal:2008}. All this
suggested that supernovae are the best candidate and triggered lot of
investigations to understand neutrino-driven winds. However, the most
recent and detailed studies indicate that neutrino-driven winds do not
get enough neutron-rich conditions to produce elements up to uranium
\cite{arcones.janka.scheck:2007, Arcones.MartinezPinedo:2011,
  Huedepohl.etal:2010, Fischer.etal:2010, Roberts.etal:2010}.

Even if not every supernova produces r-process elements,
neutrino-driven winds remain an exciting astrophysical site for the
nucleosynthesis of elements beyond iron and up to silver. First
nucleosynthesis calculations based on hydrodynamic simulations
\cite{Woosley.Hoffman:1992} showed that the seed nuclei (that capture
neutrons during the r-process) in neutrino-driven winds are heavier
than iron-group elements. The nucleosynthesis in the wind starts with
an alpha-rich freeze-out and continues with charged particle reactions
and neutron captures. When the temperature drops only the latter can
continue towards heavy nuclei via the r-process
\cite{Woosley.Hoffman:1992}. However, the charged particle reactions
or alpha process already synthesizes elements up to $N=50$ and $A=90$.

There are at least two indications that further studies are necessary
to understand the origin of elements between Sr and Ag. First,
observations of very old starts \cite{Sneden.etal:2008} indicate that
there is more than one r-process: one produces a robust pattern and
gets up to uranium, another one contributes only to the lighter heavy
elements (Sr to Ag). Qian \& Wasserburg showed that neutrino driven
winds can be the site producing these elements
\cite{Qian.Wasserburg:2001, Qian.Wasserburg:2007,
  Qian.Wasserburg:2008}. Moreover, the solar system abundances for
these lighter heavy elements cannot be explained only with the r- and
s-process as shown by \cite{Travaglio.Gallino.ea:2004,
  Montes.etal:2007}. The missing contribution can correspond to
charged particle reactions in the neutrino-driven winds
\cite{Arcones.Montes:2011, Farouqi.etal:2010} but also to fast
rotating stars at low metallicities (see \cite{Frischknecht.etal:2012}
and references therein). The process that contributes to the
production of elements between Sr and Ag can be the LEPP (lighter
element primary process \cite{Travaglio.Gallino.ea:2004}), a low
metallicity s-process \cite{Frischknecht.etal:2012} or charged
particle reactions \cite{Qian.Wasserburg:2001} combined with a weak
r-process \cite{Truran.Cowan:2000} and/or the $\nu p$-process
\cite{Pruet.Hoffman.ea:2006, Froehlich06, Wanajo:2006}.

The elements between Sr and Ag in neutrino-driven winds can be used to
understand better neutrino-driven winds and supernovae, if we combine
nucleosynthesis calculations (including the most recent experimental
data) and observations of very old stars. In contrast to the r-process
(for which many questions remain open concerning the astrophysical
site and the nuclear properties of the nuclei involved) for the
production of lighter heavy elements in neutrino-driven winds we have
(or will have soon) enough information to explain their origin and use
them to learn about the astrophysical conditions. There are three
reasons for this positive statement: 1) the required astrophysical
conditions to produce elements up to Ag are found in current
hydrodynamic simulations, 2) the still unmeasured experimental data
necessary for the nucleosynthesis calculations will be accessible in
new radioactive beam facilities because most of the nuclei involved
are relatively close to stability, 3) present and planned surveys are
rapidly improving in quality and quantity allowing to collect more
information about very old stars that provide an unique fingerprint of
early nucleosynthesis. All these advances required a parallel effort
with nucleosynthesis studies.  In this paper we use a trajectory from
\cite{arcones.janka.scheck:2007} and varying the wind parameters to
study the production of elements between Sr and Ag. This was analyzed
in previous works based on a full-consistent supernova simulation
\cite{Wanajo.Janka.Mueller:2011}, full parametric high-entropy wind
models \cite{Farouqi.etal:2010}, and simplified wind simulations
\cite{Roberts.etal:2010, Arcones.Montes:2011}. Our work complement
these previous studies giving a broader and more systematic overview
of the production of lighter heavy elements and wind parameters.

In this paper, we briefly review the nucleosynthesis occurring in
neutrino-driven winds in Sect.~\ref{sec:windnuc} and we introduce the
trajectory and the nucleosynthesis network in
Sect.~\ref{sec:method}. The results are divided into neutron-rich
winds (Sect.~\ref{sec:nrich}) and proton-rich winds
(Sect.~\ref{sec:prich}). The summary and conclusions are in
Sect.~\ref{sec:summary}.

\section{Wind nucleosynthesis}
\label{sec:windnuc}
The nucleosynthesis processes occurring in neutrino-driven winds are
primary processes, i.e., they start from neutrons and protons and not
from existing seed nuclei. In the outer layers of the neutron star
temperature and density are high and the composition is dominated by
nucleons, although light cluster ($^2$H, $^3$H, $^3$He, $^4$He) can
also be present (see
e.g.~\cite{arcones.matinezpinedo.etal:2008}). This matter gets ejected
after absorbing enough energy from neutrinos that are emitted during
the cooling of the neutron
star~\cite{duncan.shapiro.wasserman:1986}. As matter is ejected its
temperature and density drop and this allows nuclear reactions to
start building alpha particles and seed nuclei. The composition of the
seed distribution and the subsequent evolution towards synthesizing
elements beyond the iron group depend on three wind quantities (see
e.g.,~\cite{Meyer:1993, Qian.Woosley:1996, hoffman.woosley.qian:1997,
  Thompson.Burrows.Meyer:2001}): entropy, expansion time scale, and
electron fraction. These parameters determine the ratio between
nucleons and seed nuclei: neutron-to-seed ($Y_n/Y_{\mathrm{seed}}$)
and proton-to-seed ($Y_p/Y_{\mathrm{seed}}$) ratios. Depending on the
value of these ratios at temperatures around 3~GK, three processes are
possible: 1) r-process \cite{hoffman.woosley.qian:1997,
  Freiburghaus.Rembges.ea:1999, Thompson.Burrows.Meyer:2001} if
$Y_n/Y_{\mathrm{seed}}\gtrsim 100$, 2) weak r-process
\cite{Truran.Cowan:2000} if $Y_n/Y_{\mathrm{seed}}\sim 1$, and 3) $\nu
p$-process \cite{Pruet.Hoffman.ea:2006, Froehlich06, Wanajo:2006} if
$Y_n/Y_{\mathrm{seed}}$ is very small and $Y_p>Y_n$, i.e., proton-rich
conditions.

How the wind parameters affect the neutron-to-seed ratio has been
extensively studied in the context of the r-process, see
e.g.~\cite{Meyer:1993, hoffman.woosley.qian:1997}. The \emph{entropy}
in radiation-dominated environments depends on temperature and
density: $S \propto T^3/\rho$. Therefore, high entropy is achieved at
high temperatures and low densities and favors high neutron-to-seed
ratios. High temperature implies energetic photons that destroy the
seed nuclei into nucleons, thus reducing $Y_{\mathrm{seed}}$ and
increasing $Y_n$. In addition the formation of seed nuclei starts with
three body reactions: 3$\alpha \rightarrow ^{12}$C and $^{4}$He
$(\alpha n, \gamma)^9$Be $(\alpha, n)^{12}$C. At low density the
probability that these reactions occur is small and this results in
lower $Y_{\mathrm{seed}}$. The \emph{expansion time scale} indicates
how fast matter expands during the alpha and seed formation at
temperature around 0.5~MeV: $\tau = \frac{r}{v}|_{T=0.5 \mathrm{MeV}}$
\cite{Qian.Woosley:1996}, with $r$ and $v$ being radius and velocity,
respectively. If the expansion is very fast (i.e., small $\tau$), the
three body reactions do not have enough time to synthesize seed nuclei
and $Y_{\mathrm{seed}}$ stays low. The \emph{electron fraction} or
electron abundance tells us if the wind is neutron rich ($Y_e<0.5$) or
proton rich ($Y_e>0.5$). Since the environment is charge neutral the
abundance of electrons is equal to the abundance of protons (free and in
bound nuclei): $Y_e = \sum_i Z_i Y_i$ with $Z_i$ the proton number of
nucleus $i$ and $Y_i$ its abundance. In order to calculate the wind
nucleosynthesis one needs the electron fraction at high temperatures
($T\sim 10$~GK) when the matter is still in nuclear statistical
equilibrium and the composition consists mainly of neutrons and
protons. Under such conditions $Y_e = Y_p/(Y_p+Y_n)$. The subsequent
evolution of $Y_e$ is done within the nucleosynthesis network
(Sect.~\ref{sec:method}) where weak reactions are included. If the
electron fraction is very low $Y_e \lesssim 0.3$ the neutron-to-seed
ratio will be high enough for the r-process to occur. If $0.4 \lesssim
Y_e \lesssim 0.5$ only a weak r-process is possible unless the entropy
is very high and/or the expansion time scale very short. For $Y_e >
0.5$ the $\nu p$-process occurs if enough neutrons can be produced by
antineutrino absorption on protons. Here we will ignore the r-process
because there are already numerous studies about wind parameters and
r-process and current hydrodynamic supernova simulations show that
extreme neutron-rich conditions are not found in general (see
\cite{arnould.goriely.takahashi:2007, arcones.thielemann:2013} and
references therein).  Our focus here is on the weak r-process
(Sect.~\ref{sec:nrich}) and the $\nu p$-process
(Sect.~\ref{sec:prich}), both can occur based on current models and
produce heavy elements between Sr and Ag.

\section{Astrophysical conditions and nucleosynthesis network}
\label{sec:method}

The evolution of wind trajectories varies for different progenitors,
explosion energy, anisotropic evolution of the explosion, and
during time  after the explosion (see e.g.,
\cite{arcones.janka.scheck:2007, Arcones.Janka:2011,
  Fischer.etal:2010, Roberts.etal:2010}). Therefore, a complete
picture of the nucleosynthesis from neutrino-driven winds requires a
large number of three-dimensional simulations following the explosion
and the wind evolution during few seconds and for different
progenitors. This is far from being possible because of the huge
computational time necessary and also due to the still uncertain
details of the explosion \cite{Janka:2012, burrows:2013} and wind
evolution \cite{arcones.thielemann:2013}. Among these uncertainties
are the neutrino matter interactions on the surface of the neutron
star \cite{Roberts:2012, Roberts.Reddy:2012,
  MartinezPinedo.etal:2012}, neutrino oscillations
\cite{McLaughlin.etal:1999, Duan.etal:2011}, rotation and magnetic
fields \cite{Thompson.Duncan:1993, Thompson:2003,
  Metzger.etal:2007}. In view of these open questions and
uncertainties our strategy here is to explore the impact of the wind
parameters (Sect~\ref{sec:windnuc}) using a single trajectory and
varying its entropy, expansion timescale and initial electron
fraction. This cannot determine the integrated nucleosynthesis from
neutrino-driven winds but helps to understand the dependencies of
abundances and wind parameters as well as to suggest possible
constraints based on observations (see e.g. \cite{Arcones.Montes:2011,
  wanajo:2013}) and chemical evolution \cite{hansen.etal:2013}.

\begin{figure}[!h]
\begin{center}
\includegraphics[width=0.5\linewidth]{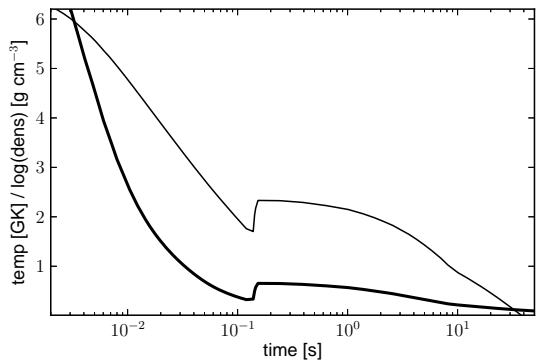}
\caption{Evolution of the temperature (thin line) and the logarithmic
  of the density (thick line) for the trajectory ejected 5~s after
  bounce of the model M15l1r1 \cite{arcones.janka.scheck:2007}. The
  reverse shock produces the jump between 0.1~s and 0.2~s.}
\label{fig:traj}
\end{center}
\end{figure}

The trajectory used in our investigation corresponds to the explosion
model M15l1r1 presented in \cite{arcones.janka.scheck:2007} and it is
ejected 5~s after bounce. This simulation has a simple but very
efficient neutrino transport that allows to study the evolution of the
wind for various progenitors in one and two dimensions
\cite{arcones.janka.scheck:2007, Arcones.Janka:2011}. The evolutions
of the density and temperature for the trajectory are shown in Fig.~\ref{fig:traj}. In
the neutrino-driven wind, matter expands very fast as indicated by the
rapid initial drop of density and temperature ($t\lesssim0.1$~s). The
wind moves through and collides with the early, slow-moving supernova
ejecta. The result is a reverse shock where kinetic energy is
transformed into internal energy producing an increase of temperature
and density (see \cite{arcones.janka.scheck:2007} for more details
about the reverse shock). The evolution after the reverse shock is
significantly slower and matter can maintain almost constant
thermodynamic conditions during half a second or even longer. However,
the details of the evolution after the reverse shock strongly
depend on the pressure distribution of the slow supernova ejecta and
this is affected by multidimensional evolution during the explosion
(see \cite{Arcones.Janka:2011} for more details).

The trajectory has the following wind parameters:
$S_{\mathrm{wind}}=78 k_{\mathrm{B}}/\mathrm{nuc}$, $\tau = 5$~ms, and
$Y_e=0.484$. These values are within the typical values obtained in
different hydrodynamic simulations \cite{arcones.janka.scheck:2007,
  Wanajo.Nomoto.ea:2009, Arcones.Janka:2011, Fischer.etal:2010,
  Roberts.etal:2010, Roberts:2012}. We employ this trajectory as
reference case and vary the parameters to cover typical conditions
values found in wind simulations. The entropy is decreased and
increased by a 30\% by changing the density as in a radiation
dominated environment, $S\propto T^3/\rho$. The time scale is modified
from 1~ms to 10~ms by using a factor over all the trajectory
time. Finally the initial electron fraction is given as a network
parameter and it is necessary to calculate the initial NSE
composition. The following evolution of $Y_e$ is performed within the
network where we include neutrino reactions on nucleons. This implies
that the neutrino luminosities and energies (which are given also as
network parameters) should be consistent with the electron fraction.

 Each network calculation starts at a temperature of $\sim 10$~GK,
 thus assuming nuclear statistical equilibrium composition for a given
 initial Ye. The evolution of the composition is followed using a full
 reaction network \cite{Froehlich06}, which includes 4053 nuclei from
 H to Hf including both neutron- and proton-rich isotopes. Theoretical
 reactions rates are from the statistical code NON-SMOKER
 \cite{Rauscher.Thielemann:2000} and experimental rates are included
 \cite{Angulo.Arnould.ea:1999} when available. The theoretical weak
 interaction rates are the same as in \cite{Froehlich06} and we use
 experimental beta-decay rates when available \cite{NuDat2}.

\section{Neutron-rich winds}
\label{sec:nrich}

\begin{figure}[!h]
\begin{center}
\includegraphics[width=0.5\linewidth]{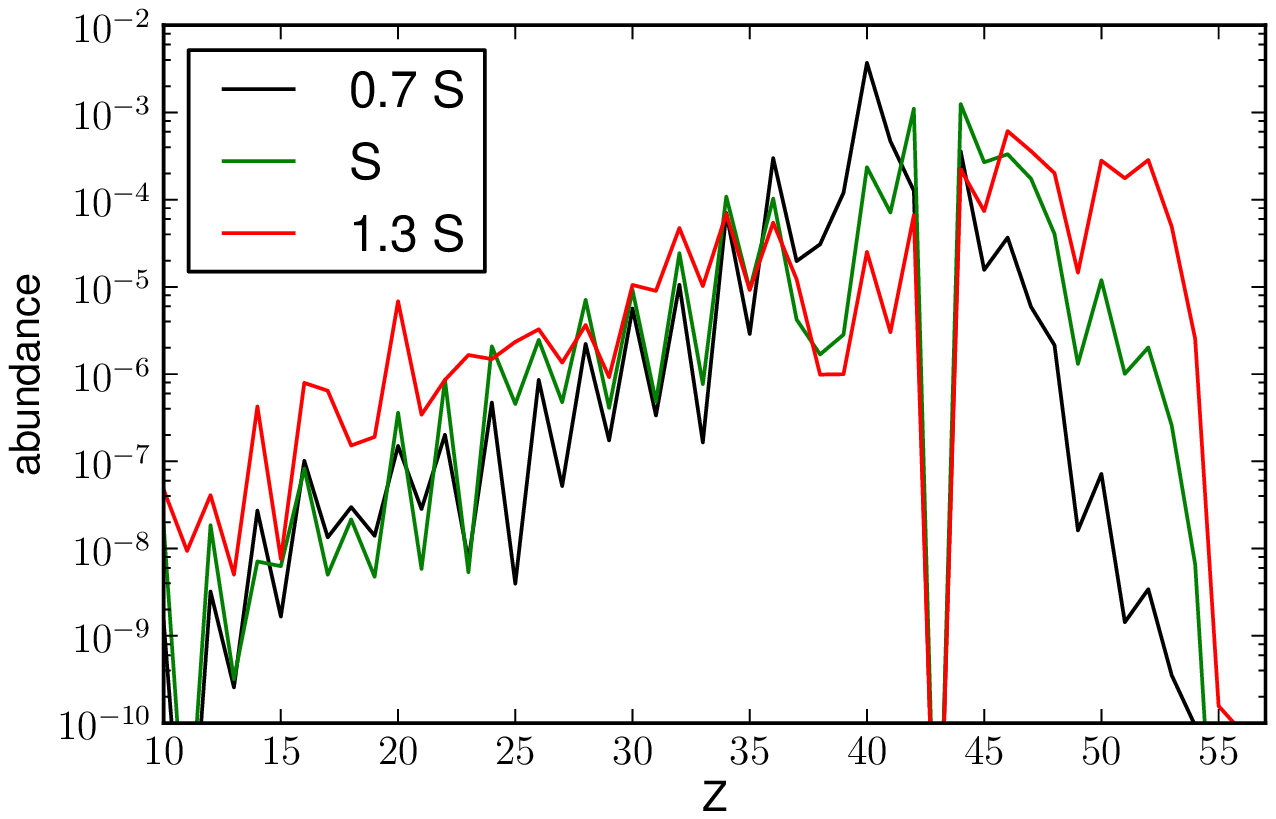}\\
\includegraphics[width=0.5\linewidth]{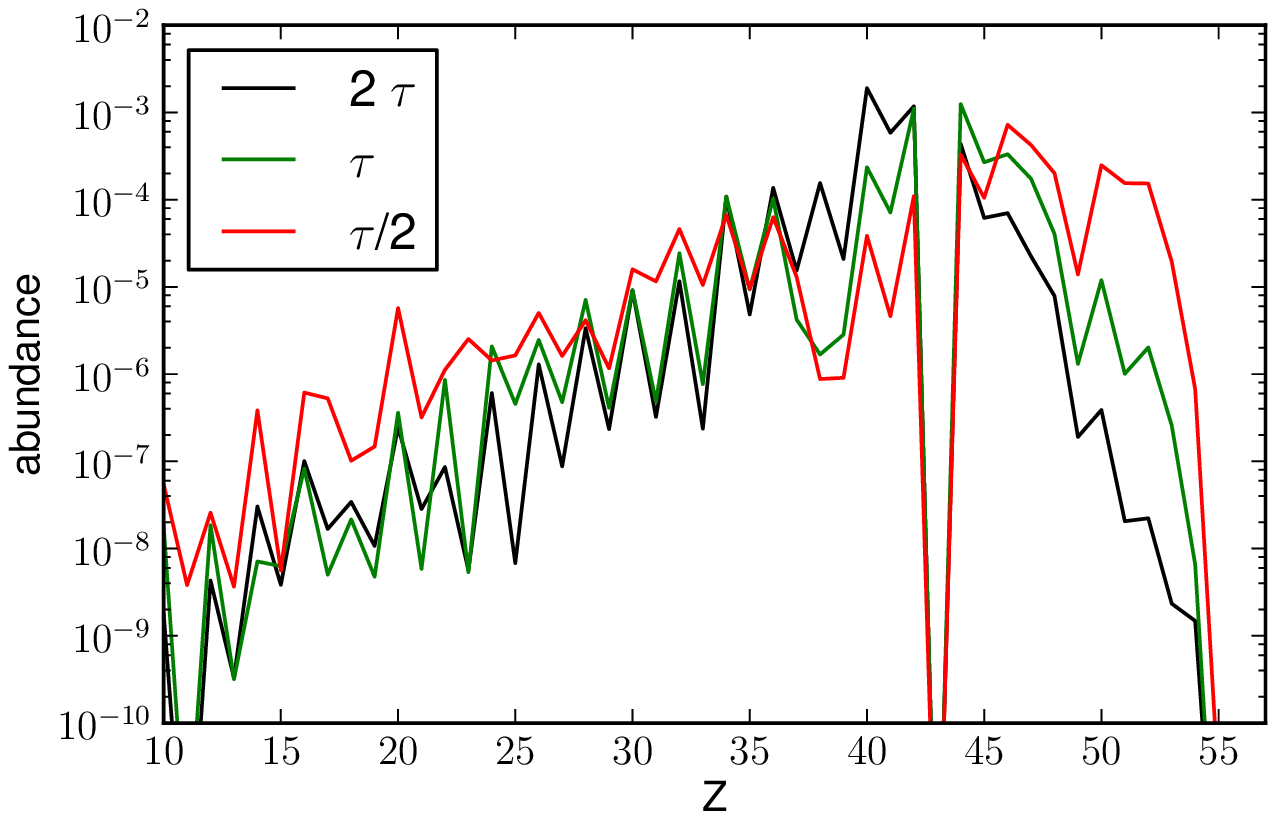}\\
\includegraphics[width=0.5\linewidth]{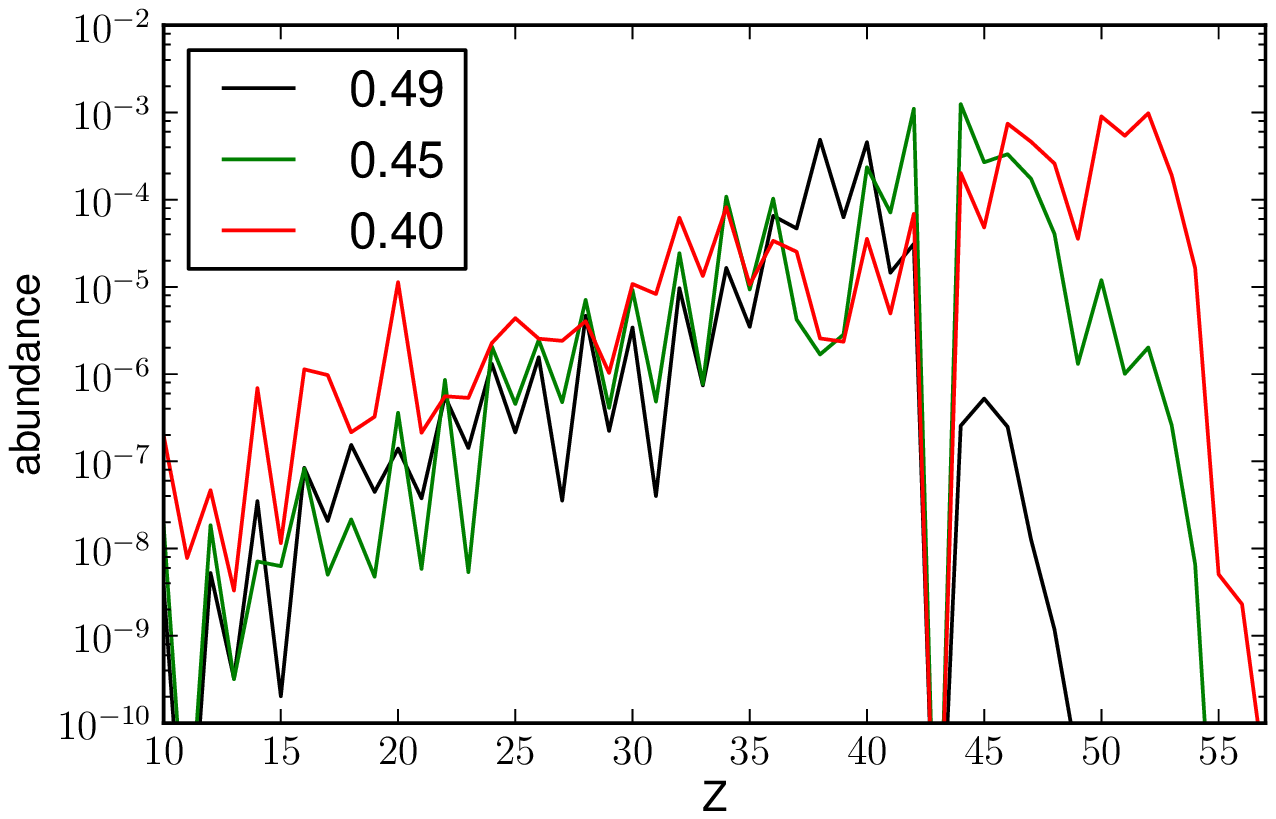}\\
\caption{Elemental abundances after decay to stability showing the
  impact of entropy, expansion time scale and electron fraction, from
  the top to the bottom panel.}
\label{fig:nrich_windpar_ab}
\end{center}
\end{figure}

In slightly neutron-rich winds a weak r-process depends on wind
parameters in the same way as the r-process occurring in very
neutron-rich winds. However, there are two main differences between
these two processes: 1) the weak r-process produces elements only
below the second peak ($Z<56$) in contrast to the r-process that can
go up to uranium, 2) in the weak r-process matter moves towards higher
Z via beta decays but mainly via $(\alpha, n)$ reactions
\cite{Woosley.Hoffman:1992}, while in the r-process beta decay is the
dominant reaction to change isotopic chain. In order to explore the
behavior of the weak r-process we have varied the wind parameters and
the resulting abundances are shown in
Fig.~\ref{fig:nrich_windpar_ab}. In general higher neutron-to-seed
ratio (i.e., higher entropy, shorter expansion time scale, lower
electron fraction) leads to heavier elements. Note that the green
line, shown in the three panels, is the reference case and corresponds
to unchanged entropy and expansion time scale and $Y_e=0.45$.

 In the top panel of Fig.~\ref{fig:nrich_windpar_ab} the entropy is
 increased (1.3~S) and decreased (0.7~S) by a 30\% while the expansion
 time scale is unmodified and $Y_e=0.45$. For the low entropy case
 there is a maximum around $Z=40$ and strong odd-even effects below
 this element. The maximum corresponds to all stable Zirconium isotopes
 with $^{90}$Zr and $^{94}$Zr being the most abundant ones. Moreover,
 only the isotope $^{88}$Sr is present in the final abundances for
 $Z=38$. These two features clearly indicate that the path does not
 moves up through stability but through unstable neutron-rich isotopes
 and eventually decays populating the stable isotopes reachable from
 the neutron-rich side. If the path would move only along stability,
 the final abundance would contain other Sr isotopes and $^{94}$Zr
 would not be very abundant. Therefore, the maximum at $Z=40$ is not
 directly linked to the shell closure at $N=50$ for these
 conditions. The path gets to the neutron rich side until $N=50$ and
 continues via beta decays and neutron captures increasing the proton
 number at constant neutron number $N=50$ until $^{86}$Kr is
 reached. This stable isotope does not beta decay allowing neutrons to
 be captured and the path to overcome the waiting point. Therefore,
 the peak around $Z=40$ is indirectly linked to the magic number
 $N=50$ in the sense that matter accumulates at $^{86}$Kr but not at
 $^{88}$Sr, $^{89}$Y, and $^{90}$Zr. Similar results were already
 discussed in \cite{Woosley.Hoffman:1992}.

 With higher entropy additional neutron captures allow matter to flow
 further away of stability and overcome the magic number $N=50$ at
 lower $Z$. This flow only stops at the shell closure $N=82$ and
 it is followed by beta decays and neutron captures, i.e. moves up in
 $Z$ at constant $N$. Therefore, the final abundances have three
 features different to the low entropy case: lower abundances in the
 region of $36<Z< 42$ (matter is moved from lower to higher $Z$),
 higher abundances for $Z>43$, and an abrupt drop at $Z\sim 55$. In
 addition, there is a small maximum in the abundances between
 $50<Z<53$ that corresponds to mass numbers $118<A<130$. This matter
 reaches $N=82$ and accumulates there before decaying to
 stability. The variations of the expansion time scale (middle panel,
 Fig.~\ref{fig:nrich_windpar_ab}) and the electron fraction (bottom panel,
 Fig.~\ref{fig:nrich_windpar_ab}) have a similar impact to the changes
 in entropy.

\begin{figure*}[!h]
\begin{center}
\includegraphics[width=0.5\linewidth]{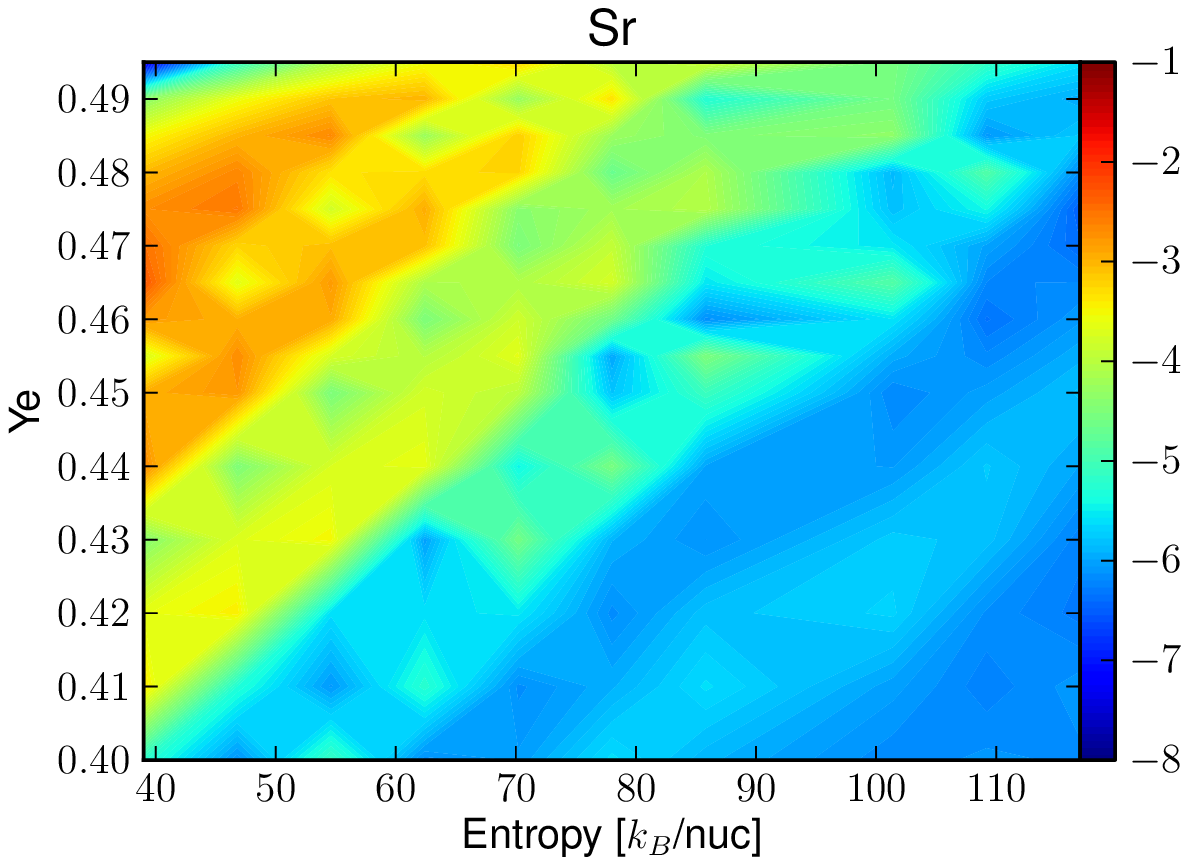}%
\includegraphics[width=0.5\linewidth]{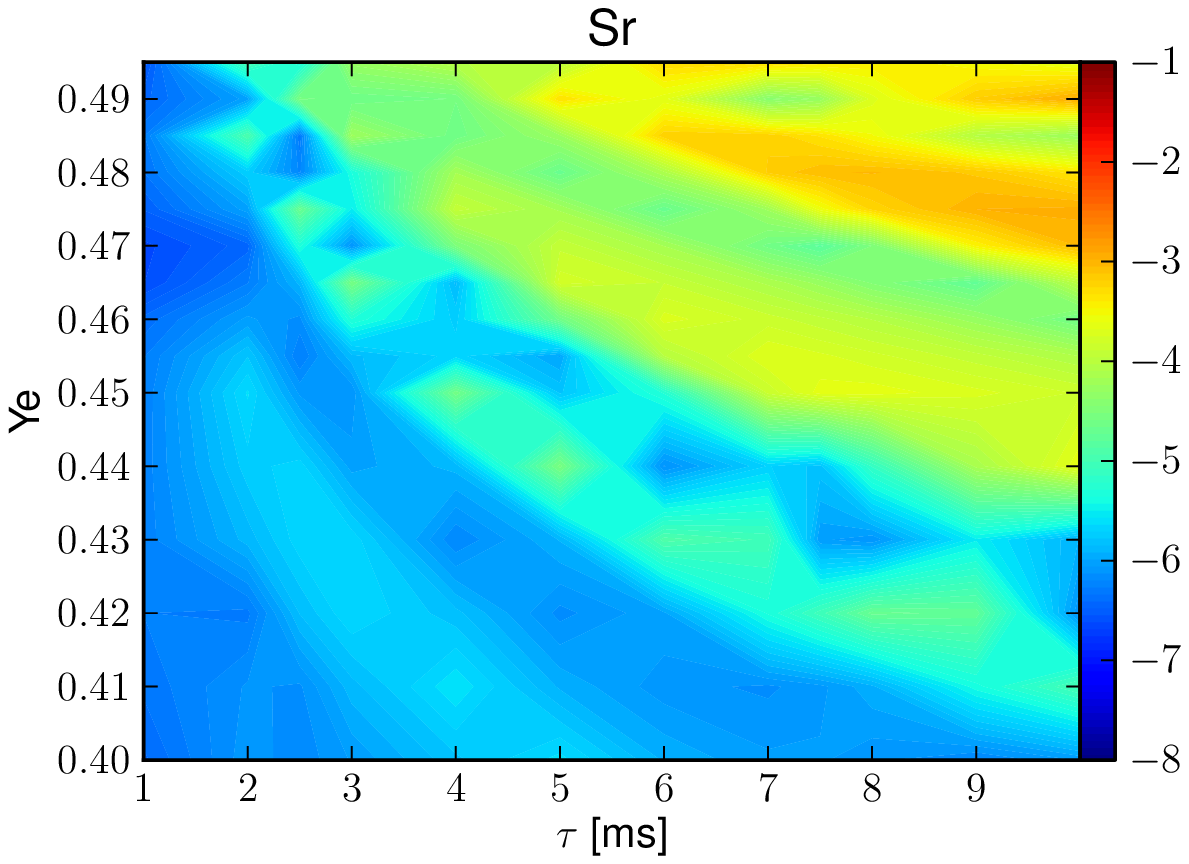}\\
\includegraphics[width=0.5\linewidth]{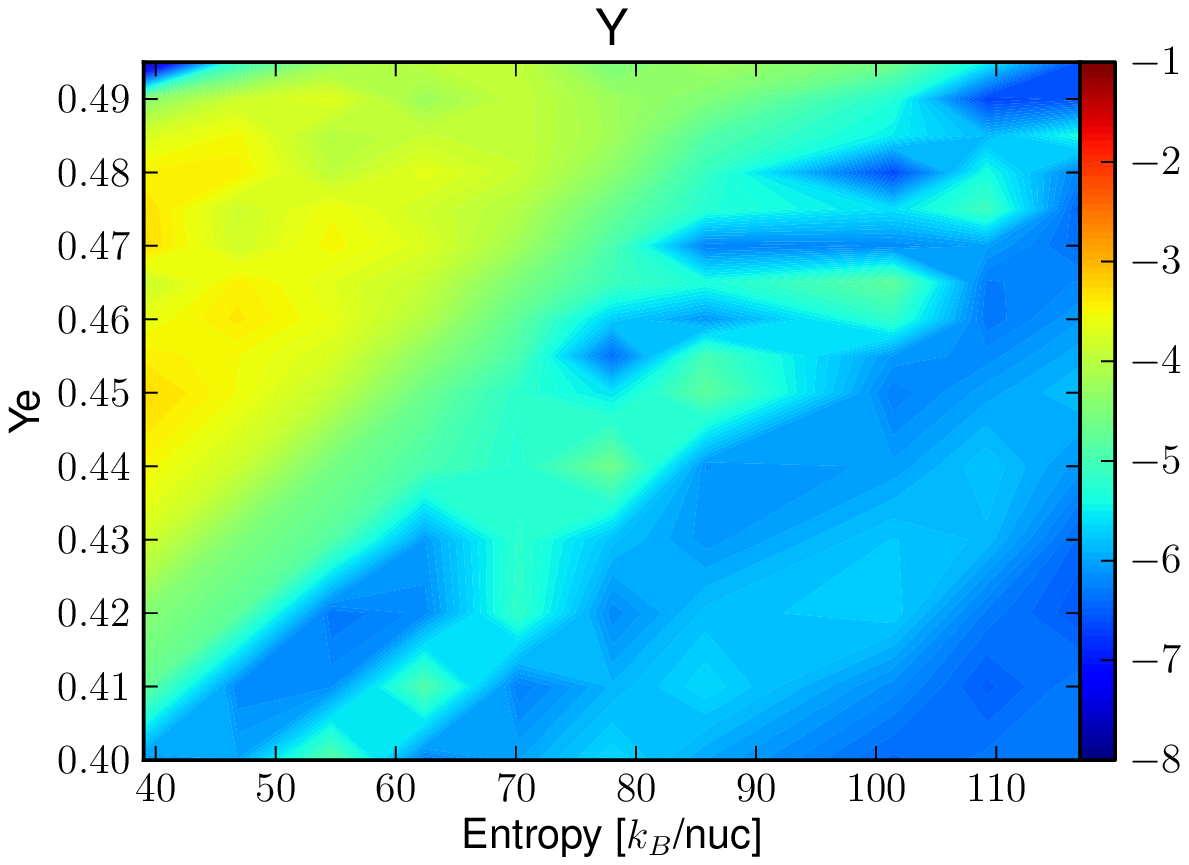}%
\includegraphics[width=0.5\linewidth]{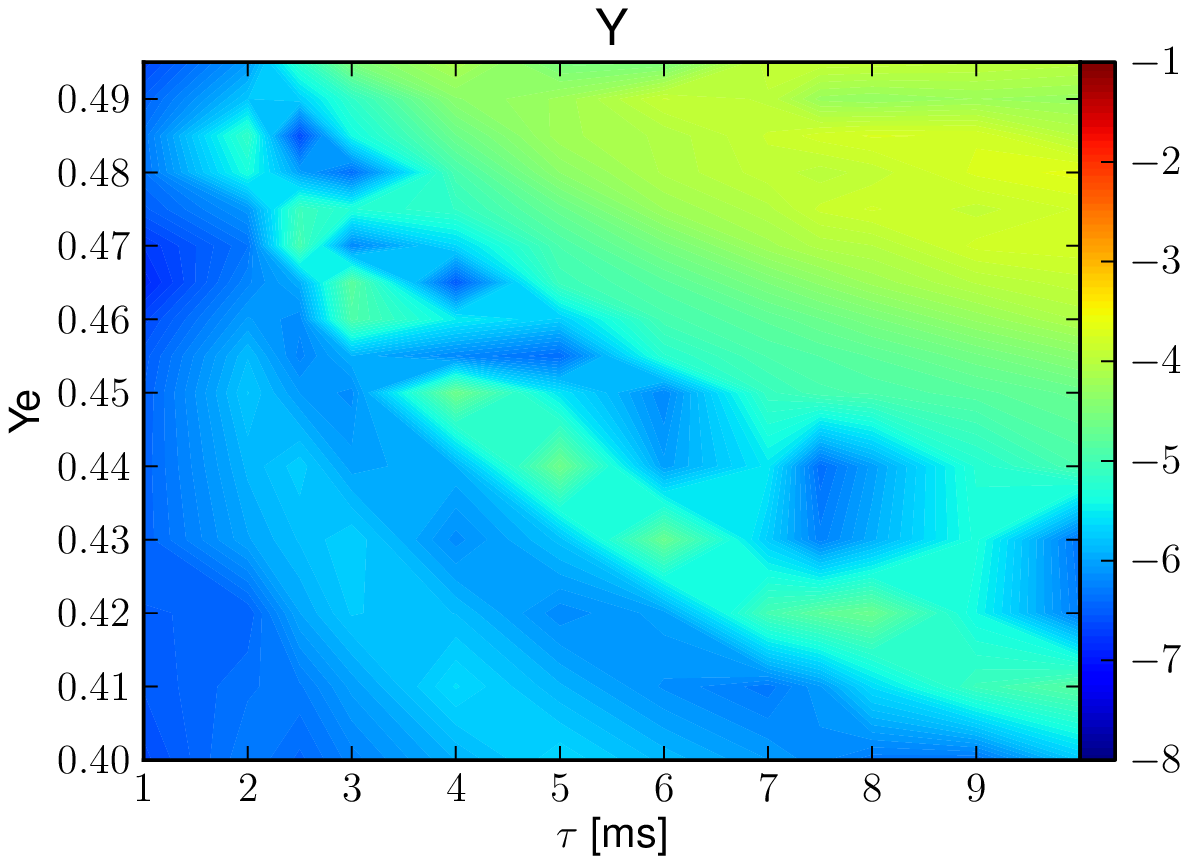}\\
\includegraphics[width=0.5\linewidth]{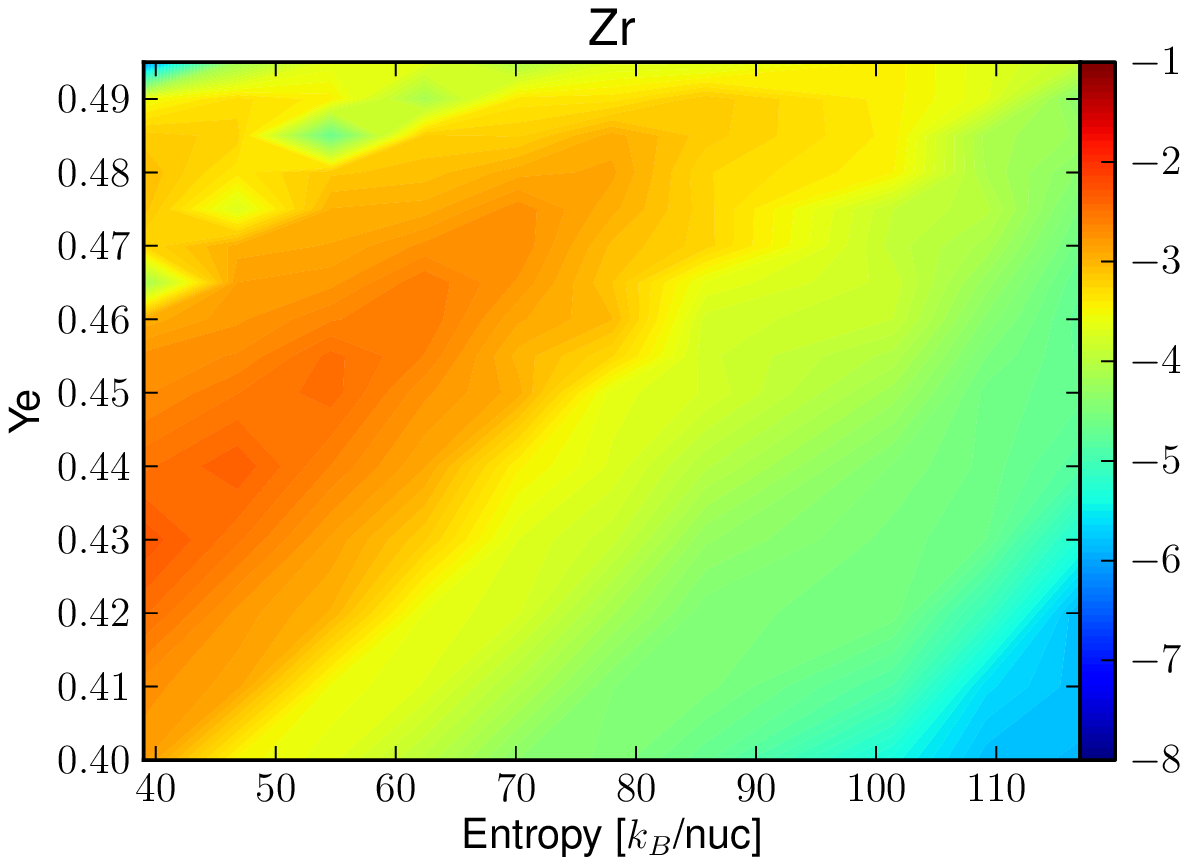}%
\includegraphics[width=0.5\linewidth]{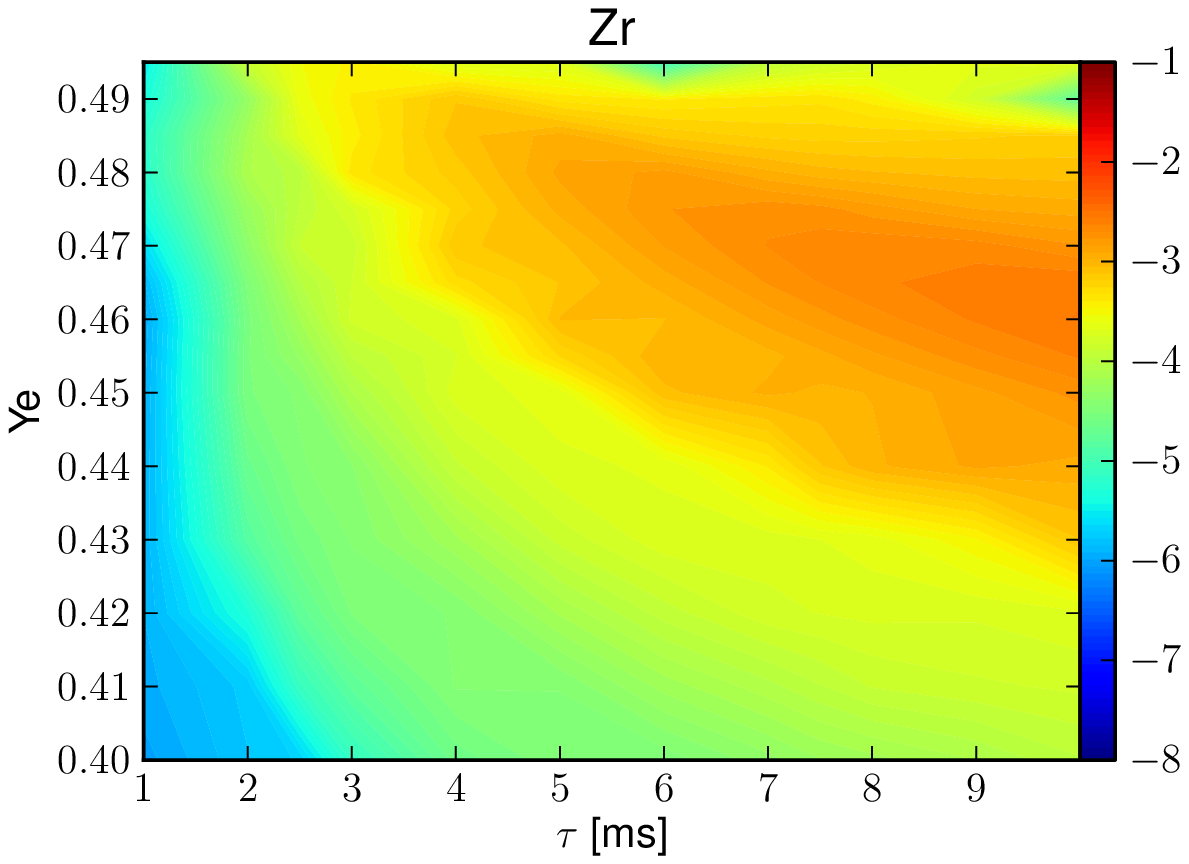}\\
\includegraphics[width=0.5\linewidth]{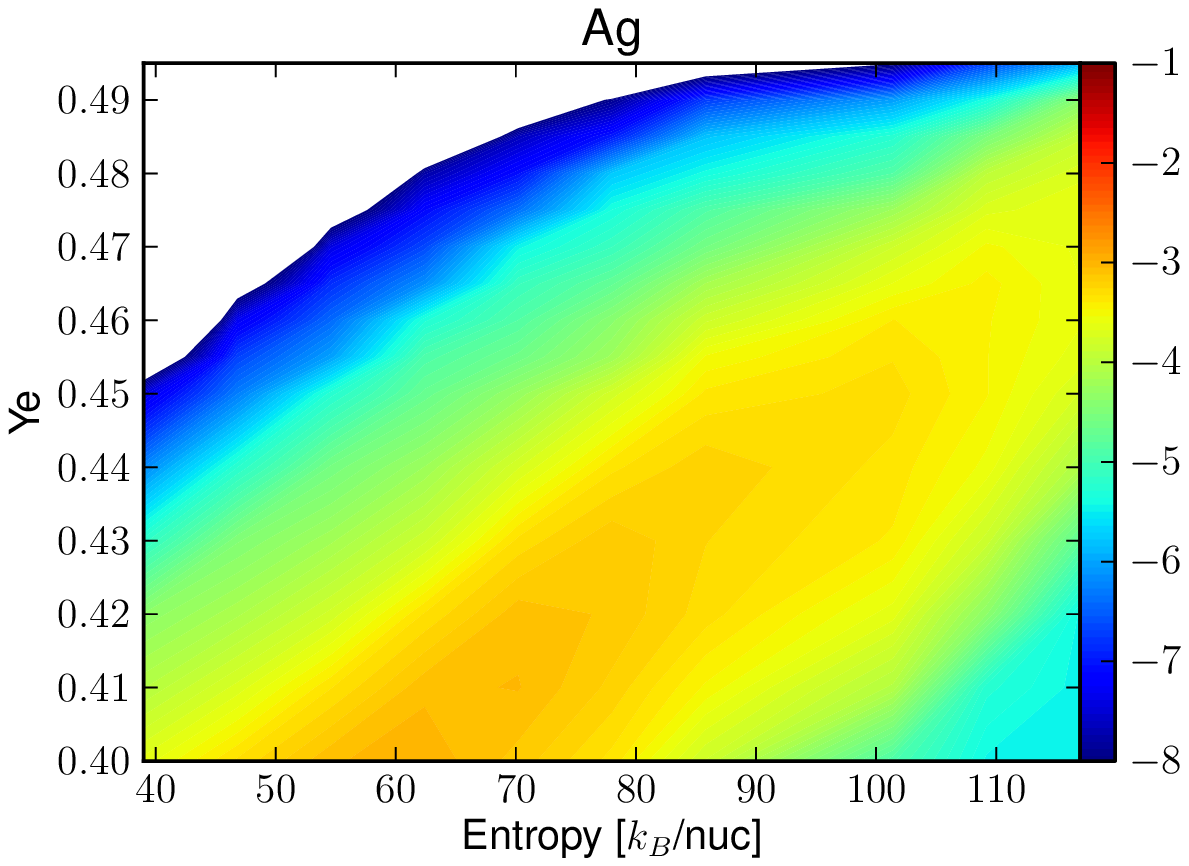}%
\includegraphics[width=0.5\linewidth]{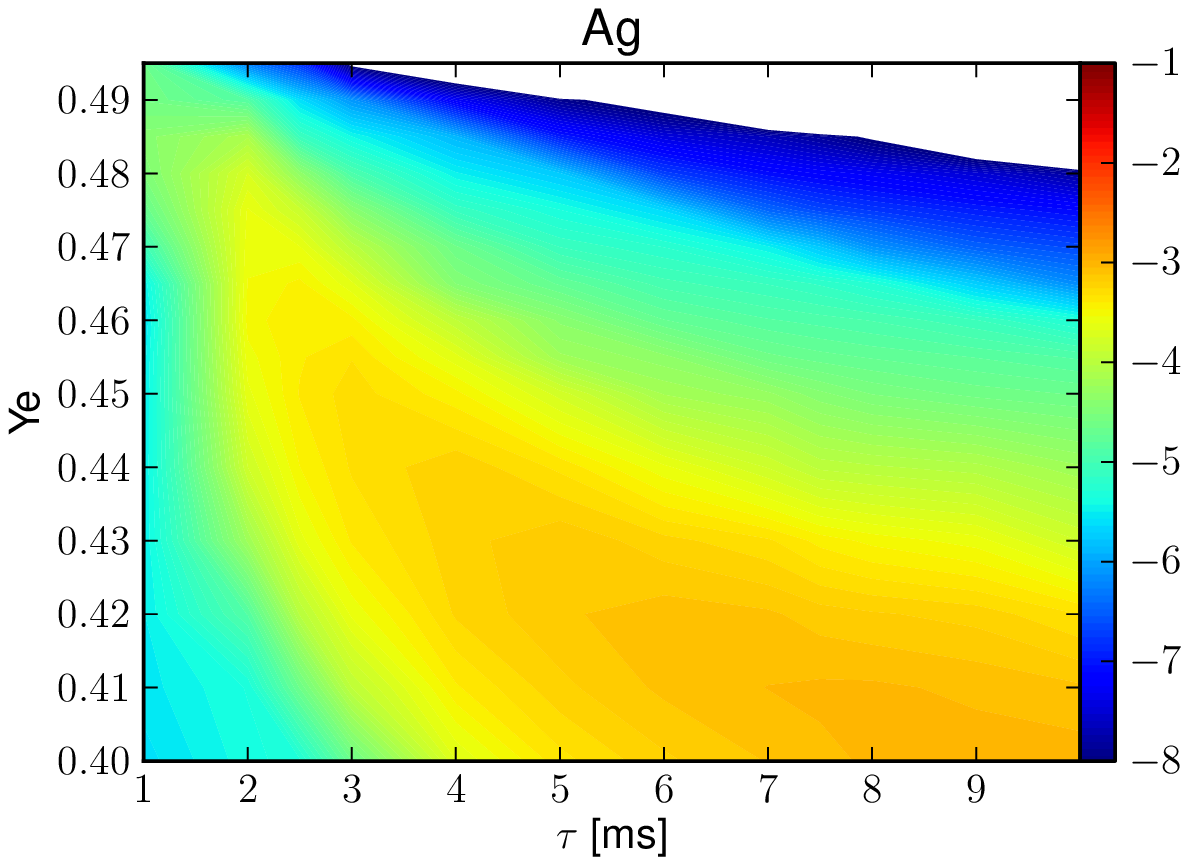}
\caption{Color contours show the abundance in log scale of Sr, Y, Zr,
  and Ag from the top to the bottom for different $Y_e$ and entropy
  (left column) or time scale (right column).}
\label{fig:nrich_elem_stau_ye}
\end{center}
\end{figure*}

Other important result is that any variation of the wind parameters
changes significantly the abundance pattern. When the neutron-to-seed
ratio increases there is not only a variation on the maximum $Z$ that
it is reached, but also the abundances are different for low $Z$. This
may be relevant if one wants to compare to observations. Here, we are
interested in the production of elements between Sr and Ag in
neutrino-driven winds. In Fig.~\ref{fig:nrich_windpar_ab} one can see
that these elements are very sensitive to the wind parameters. In
order to get a better overview, Fig.~\ref{fig:nrich_elem_stau_ye}
shows the abundances of Sr, Y, Zr, and Ag for various electron
fractions and entropies (left column) and time scales (right
column). For low neutron-to-seed ratio, i.e. at high $Y_e$, low
entropy and long time scale, Sr, Y, and Zr dominate the
composition. While for higher neutron-to-seed ratios silver is more
abundant at expenses of Sr, Y, and Zr whose abundances are lower. An
important feature are the structures in the panels for Sr, Y, and
Zr. Their abundances oscillate and present several minima and maxima
when wind parameters are varied. In contrast, the abundances of silver
smoothly change with wind parameters. Observations
\cite{Sneden.etal:2008} of the r-process indicate that the abundance
pattern between Sr and Ag is not so robust as above $Z\approx 50$ (in
stars with high enrichment of r-process). However, strong variations,
as we find here, are also not expected. This could be an indication
that the wind may not be neutron rich or that the wind parameters
stays rather constant during the wind phase. However, this implies
that all winds should have same entropy, expansion time scale, and
electron fraction and current simulations show that the wind
parameters change for different progenitors and at different times during
the wind evolution. Other possibility is that the amount of
neutron-rich material ejected is very small
\cite{Wanajo.Janka.Mueller:2011} and does not contribute to the
observed abundances.

\section{Proton-rich winds}
\label{sec:prich}

When electron neutrinos and electron antineutrinos have similar
energies the wind becomes proton-rich due to the neutron-proton mass
difference. This can happen if neutral current reactions dominate in
the region where neutrinos decouple from matter, i.e. at the
neutrinosphere \cite{Huedepohl.etal:2010, Fischer.etal:2012}. Since
the electron fraction of the wind is not yet well determined and there
are simulations pointing to both neutron- and proton-rich conditions,
we analyze now the nucleosynthesis in proton-rich winds.

\begin{figure}[!h]
\begin{center}
\includegraphics[width=0.5\linewidth]{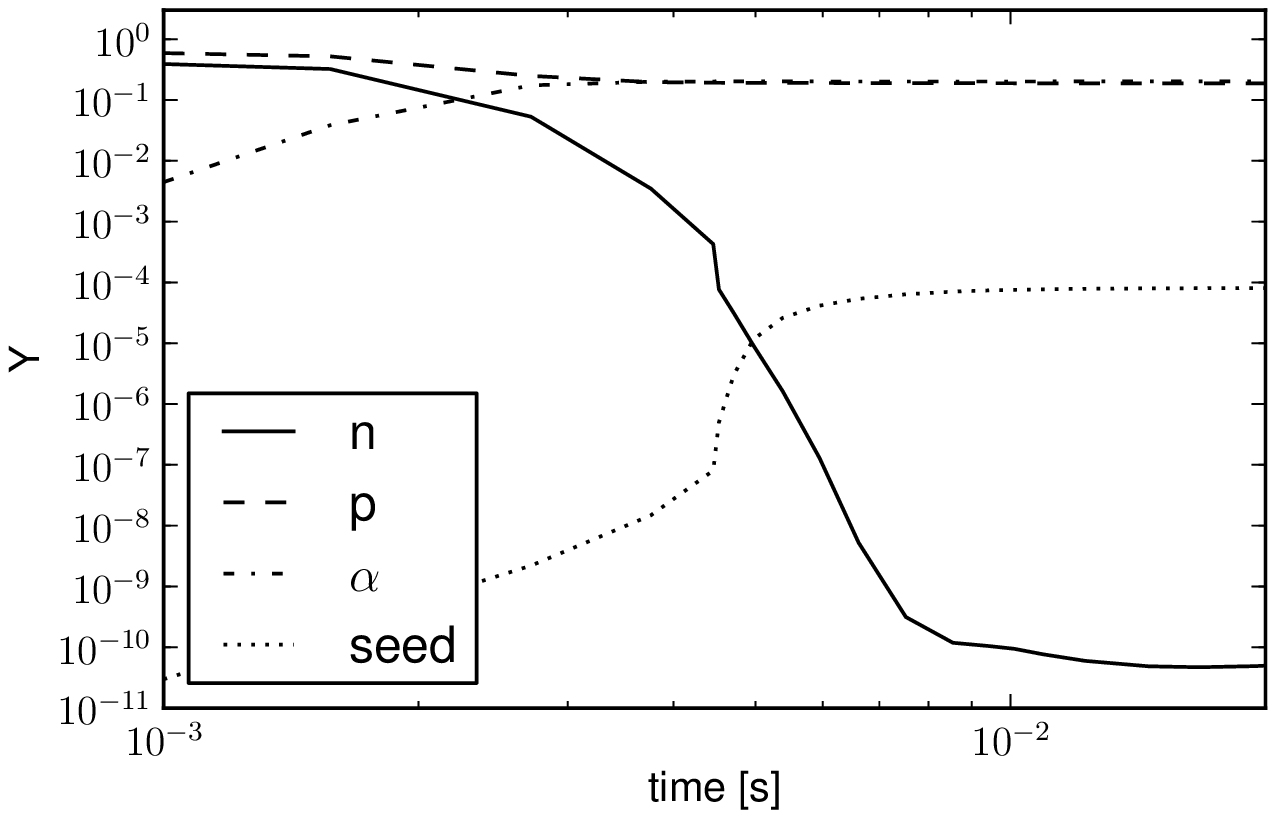}
\caption{Abundance evolution for neutrons, protons, alpha particles,
  and seed nuclei in a proton-rich wind.}
\label{fig:prich_abevol}
\end{center}
\end{figure}

In proton-rich conditions heavy elements are synthesized from seed
nuclei via charged-particle reactions, i.e., $(p,\gamma)$, $(\alpha,
\gamma)$, $(\alpha,p)$, and beta decays. These reactions allow the
flow of matter to move on the proton-rich side of stability up to
$^{64}$Ge. This nucleus has a beta half life much longer than the
expansion time scale of the wind, it is a bottleneck. However, the
matter in the wind is exposed to high neutrino fluxes and the
antineutrino absorption on protons produces neutrons that can be used
to overcome such bottlenecks by $(n,p)$ reactions. This is the $\nu
p$-process that was first discussed by \cite{Pruet.Hoffman.ea:2006,
  Froehlich06, Wanajo:2006} and that depends on the neutron-to-seed
ratio. Figure~\ref{fig:prich_abevol} shows the evolution of abundances
for neutrons, protons, alpha particles, and seed nuclei. The neutron
abundance drops to very small values but still plays a key role in the
synthesis of heavy elements via reactions such as $(n, p)$ and
$(n,\gamma)$. 

\begin{figure}[!h]
\begin{center}
\includegraphics[width=0.5\linewidth]{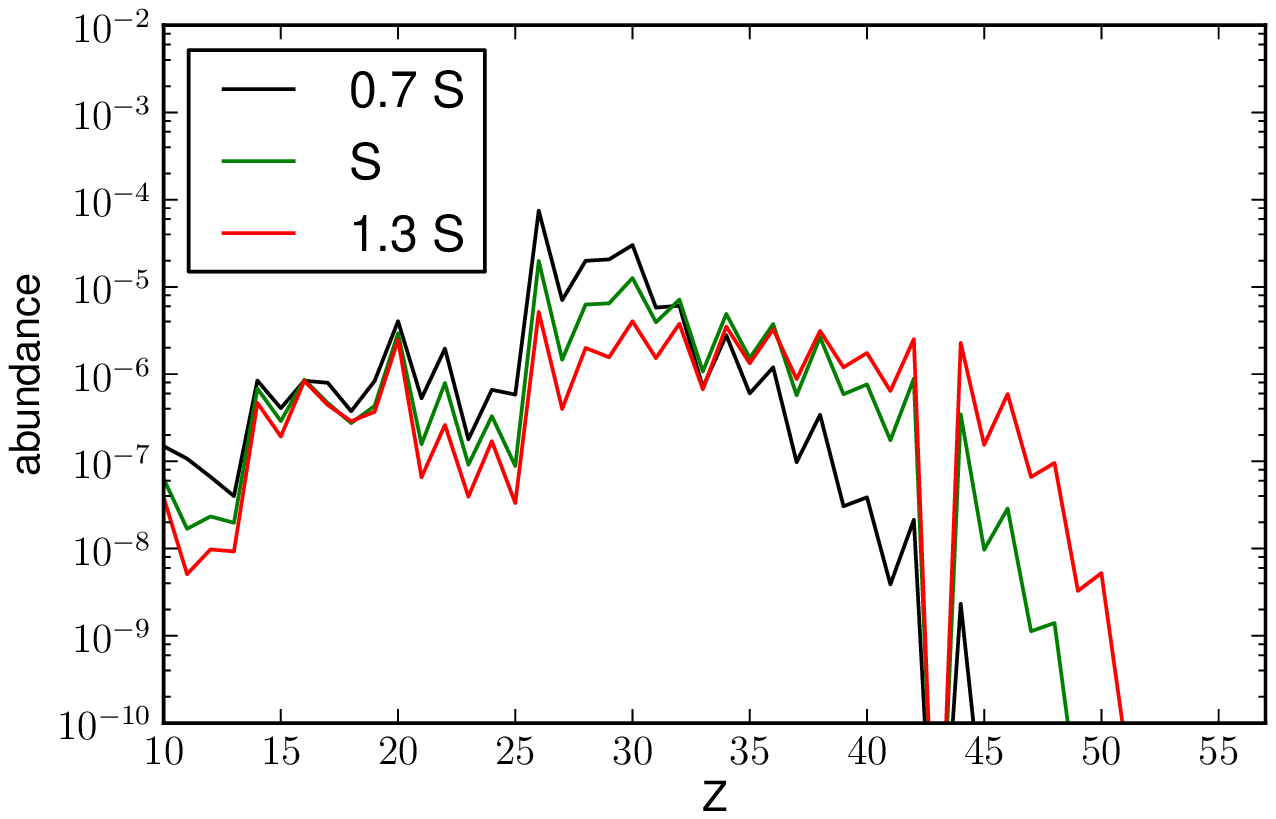}\\
\includegraphics[width=0.5\linewidth]{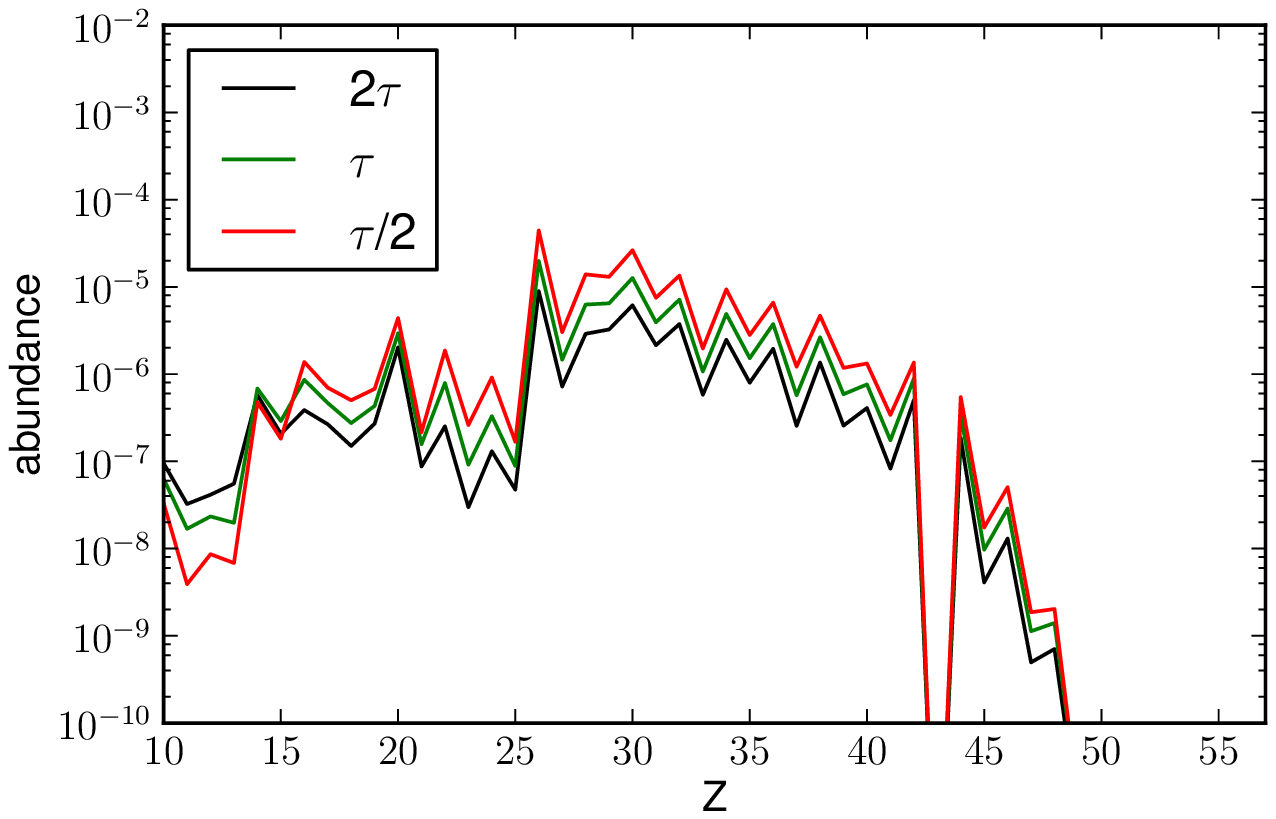}\\
\includegraphics[width=0.5\linewidth]{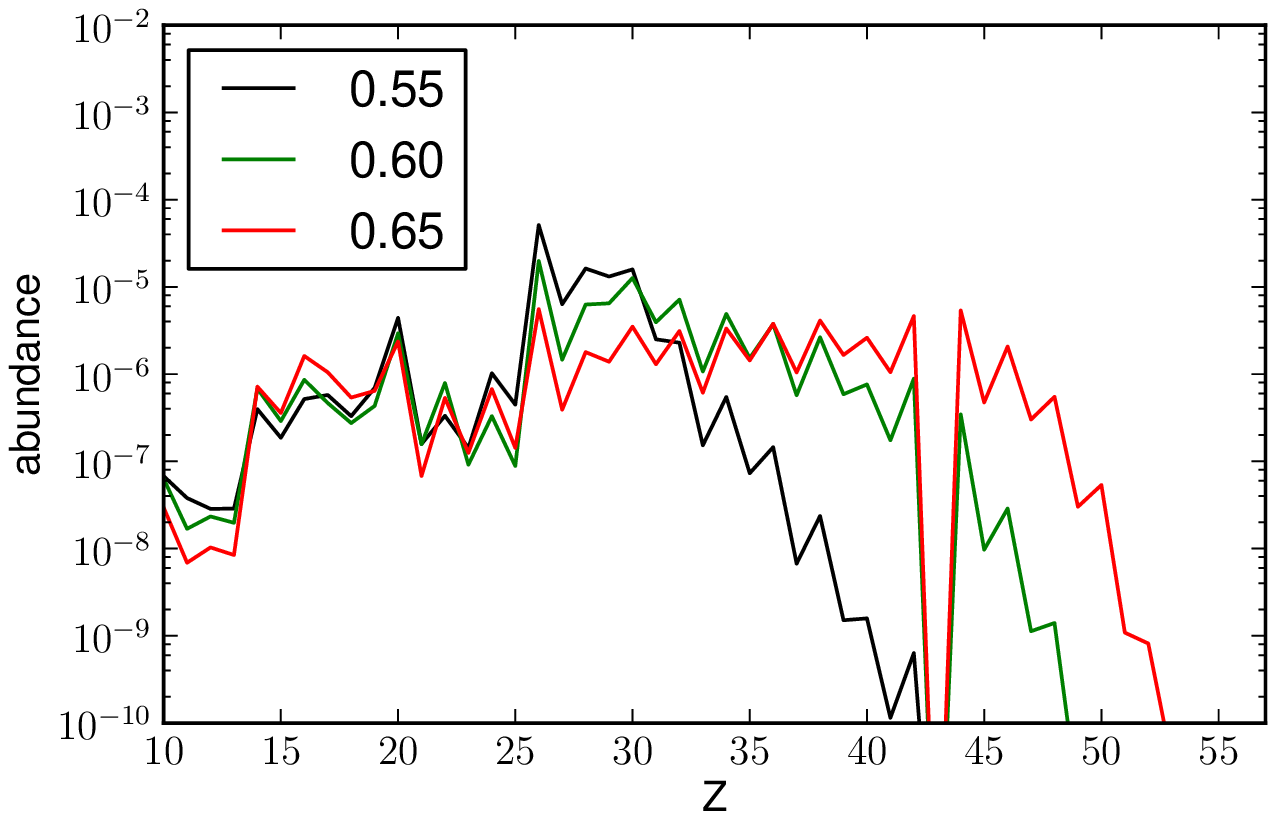}\\
\caption{Elemental abundances showing the impact of entropy, expansion
  time scale and electron fraction, from the top to the bottom panel.}
\label{fig:prich_windpar_ab}
\end{center}
\end{figure}

The impact of the wind parameters on the $\nu p$-process has been
already discussed by e.g. \cite{Pruet.Hoffman.ea:2006,
  Wanajo:2006}. Here we aim to give a more systematic overview with
focus on the production of Sr, Y, Zr, and Ag. The reactions that
control the production of heavy nuclei in proton-rich winds are the
$(n,p)$ reactions. Therefore, one needs to understand how the
neutron-to-seed ratio changes with wind parameters. The amount of
seeds ($Y_{\mathrm{seed}}$) depends on entropy and time scale in the
same way as in neutron-rich winds. Higher entropies and faster
expansions lead to lower seed and higher neutron abundances,
i.e. higher $Y_n/Y_{\mathrm{seed}}$. The impact on the abundance of
these two wind parameters and of $Y_e$ is presented in
Fig.~\ref{fig:prich_windpar_ab}.

As in neutron-rich winds, we have increased and decreased the entropy
by a 30\% while keeping the expansion time scale of the simulation and
$Y_e=0.60$ (top panel, Fig.~\ref{fig:prich_windpar_ab}). In the three
cases, Sr, Y and Zr are produced although for the wind with lower
entropy only in small amount. With higher entropy Ag and Pd can be
also synthesized and if the entropy were even higher also heavier
elements would appear. Here we keep the entropy variation within a
30\% of the original values, for more extreme variation see
\cite{Pruet.Hoffman.ea:2006}.

The expansion time scale (middle panel,
Fig.~\ref{fig:prich_windpar_ab}) has a moderate impact on the
abundances. This time scale corresponds to the expansion of matter
when alpha particles form and thus affects the seed
production. However, the expansion during other phases of the
evolution critically affects the nucleosynthesis in proton-rich
winds. Heavy nuclei are synthesized via charged particle reactions;
therefore, the time that the matter spends at high enough temperatures
(T>2GK) is also important due to the Coulomb barrier. Moreover,
neutrons are needed to overcome the bottlenecks, thus also the time
matter is exposed to neutrinos becomes relevant. This was studied in
\cite{Wanajo.etal:2011, Arcones.Frohlich:2012} by changing the
temperature of the reverse shock and the time scale of the evolution
afterwards. They found that there is an optimal temperature for the
reverse shock to form heavy nuclei. When matter is decelerated by the
reverse shock and kept at $T\sim 2$~GK, heavier elements can be
produced. If the temperature stays higher, photo-dissociation
reactions prevent the flow of matter to reach heavy elements, matter
stays in a NiCu cycle \cite{Arcones.Frohlich:2012}. If the reverse
shock is at much lower temperature or even absence (as it is the case
of low mass progenitor explosions \cite{Wanajo.Nomoto.ea:2009,
  Hoffman.Mueller.Janka:2008}), the matter expands very fast and
antineutrinos do not have enough time to produce the necessary amount
of neutrons to overcome the bottlenecks. Therefore, the expansion time
scale in proton-rich winds is important during the alpha formation and
also during the $\nu p$-process. In the middle panel of
Fig.~\ref{fig:prich_windpar_ab}, we have varied the overall evolution
by dividing or multiplying the trajectory time by a given factor. This
means that a faster expansion through the reduction of the time scale
by a factor has two effects: 1) there are less alpha particles and
seed nuclei and $Y_n/Y_{\mathrm{seed}}$ can become higher than in the
reference case, 2) matter is exposed to the electron antineutrino flux
during less time and this reduces the amount of protons that can be
converted into neutrons with the result of lower $Y_n$. Therefore, one
effect compensates the other and the final result is small changes in
the abundances for different time scales.

The electron fraction plays a very important role as it determines the
amount of free neutrons. The neutron-to-seed ratio increases when the
seed abundance decreases as we have discussed above, but also when the
neutron abundances increases. The initial electron fraction gives the
amount of proton as initially nucleons dominate the composition,
i.e., $Y_p=Y_e$ and $Y_n=1-Y_e$. When matter expands and forms seed
nuclei, neutrons are rapidly captured but also produced by antineutrino
absorption on protons. Therefore, the abundance of neutrons reaches
and equilibrium given by
\begin{equation}
  \frac{\mathrm{d}Y_n}{\mathrm{d}t} = \lambda_{\bar{\nu}_e}Y_p -
  \sum_{Z,A} N_n Y(Z,A) \langle \sigma v \rangle_{(Z,A)} =0 \, ,
\end{equation}
here $\lambda_{\bar{\nu}_e}$ is the electron antineutrino absorption
rate and $\langle \sigma v \rangle_{(Z,A)}$ is the sum of neutron capture
rates ($(n,\gamma)$ and $(n,p)$) for nucleus ($Z$,~$A$). This
expression indicates the dependence of the neutron abundances on the
proton abundances and thus on the initial electron fraction:
\begin{equation}
  N_n = \rho Y_n N_A = \frac{\lambda_{\bar{\nu}_e}Y_p
  }{\sum_{Z,A}Y(Z,A) \langle \sigma v \rangle_{(Z,A)}}\, ,
\label{eq:n-equil}
\end{equation}
where $\rho$ is the density and $N_A$ the Avogadro
number. Eq.~(\ref{eq:n-equil}) shows that neutron abundance increases
for higher $Y_p$ and $\lambda_{\bar{\nu}_e}$. The first one depends on
$Y_e$ and its effect is visible in the bottom panel of
Fig.~\ref{fig:prich_windpar_ab}: higher electron fraction
significantly increases the abundances of Sr, Y, and Zr and allows to
produce heavier elements. The antineutrino absorption rate depends on
antineutrino energy ($\epsilon_{\bar{\nu}_e}$) and number luminosity
($L_{n,\bar{\nu}_e}$): $\lambda_{\bar{\nu}_e} = L_{n,\bar{\nu}_e}
\sigma_{\bar{\nu}_ep} \propto L_{n,\bar{\nu}_e}
\epsilon_{\bar{\nu}_e}^2$. Therefore, an increase of antineutrino
luminosity and energy will have similar effect to the increase of
$Y_e$, i.e. the production of heavy elements becomes more efficient.

\begin{figure*}[!h]
\begin{center}
\includegraphics[width=0.5\linewidth]{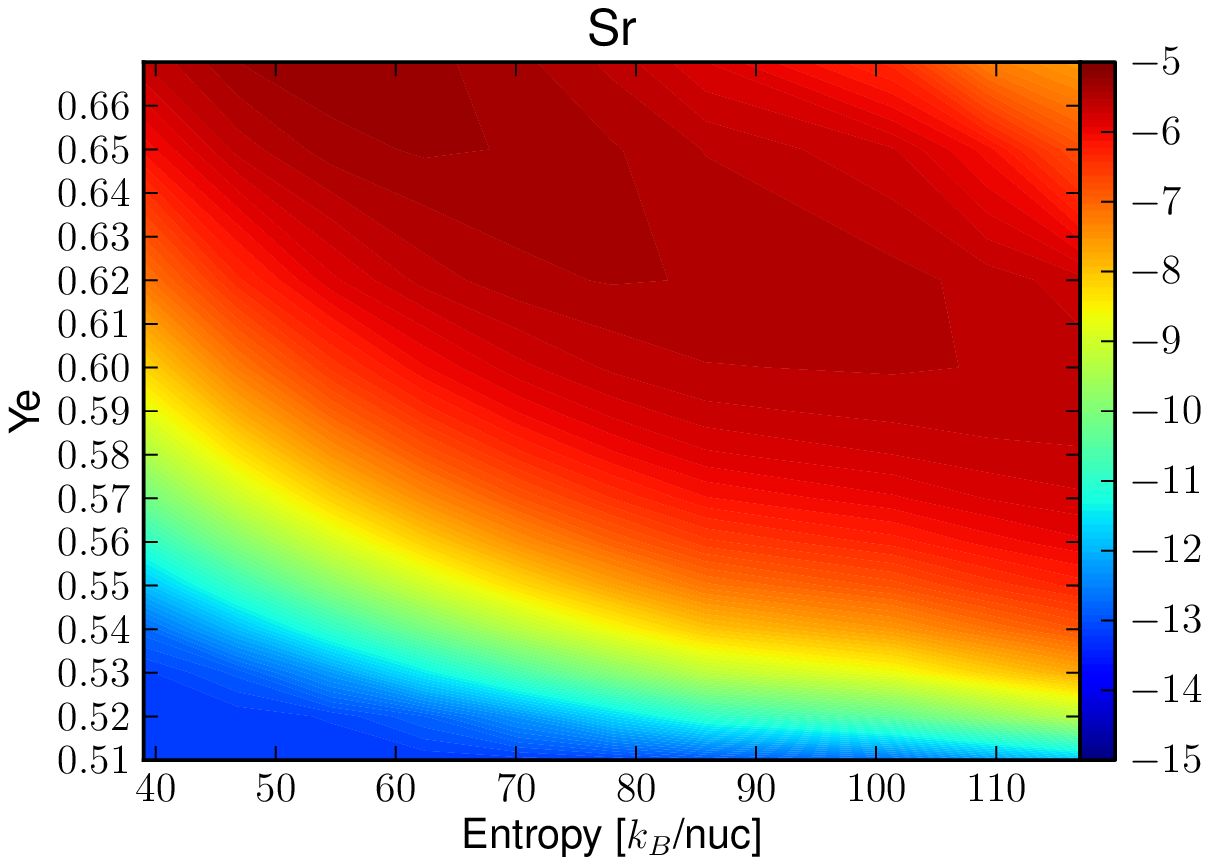}%
\includegraphics[width=0.5\linewidth]{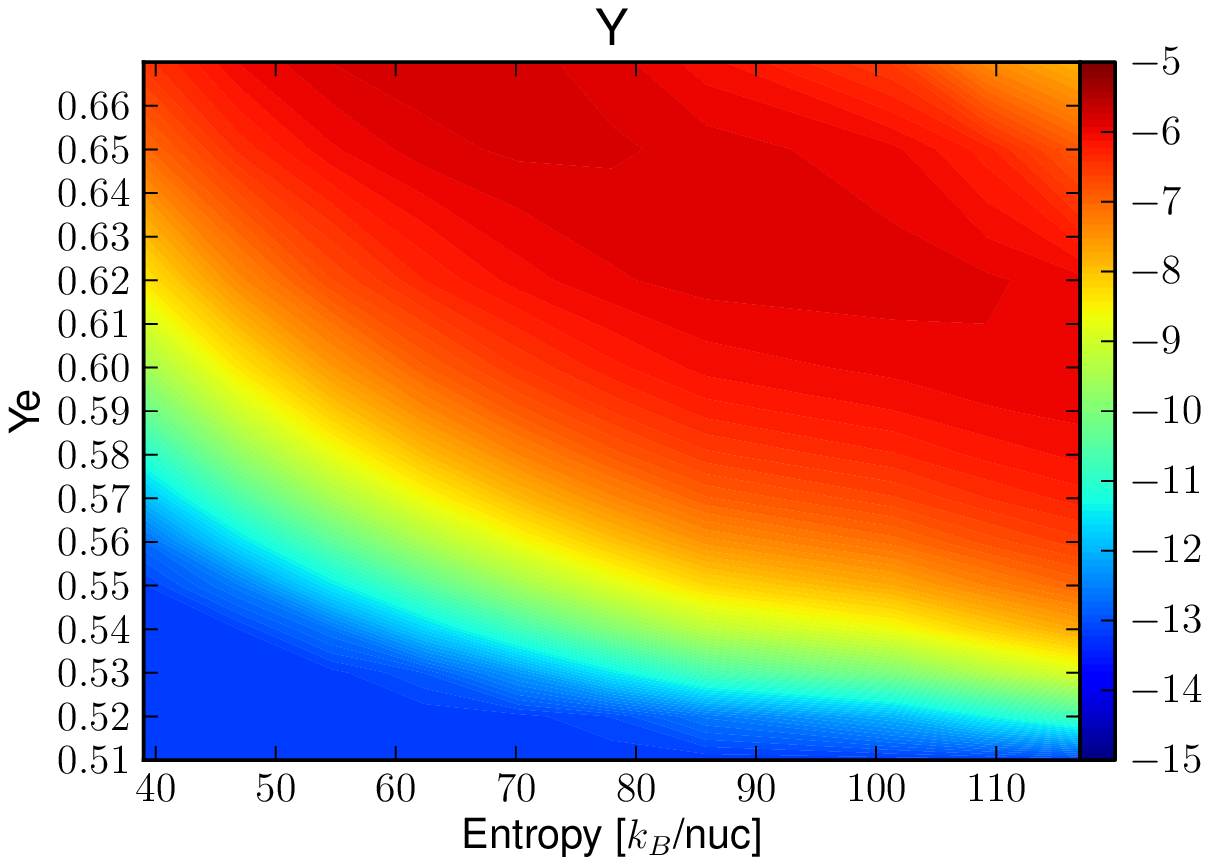}\\
\includegraphics[width=0.5\linewidth]{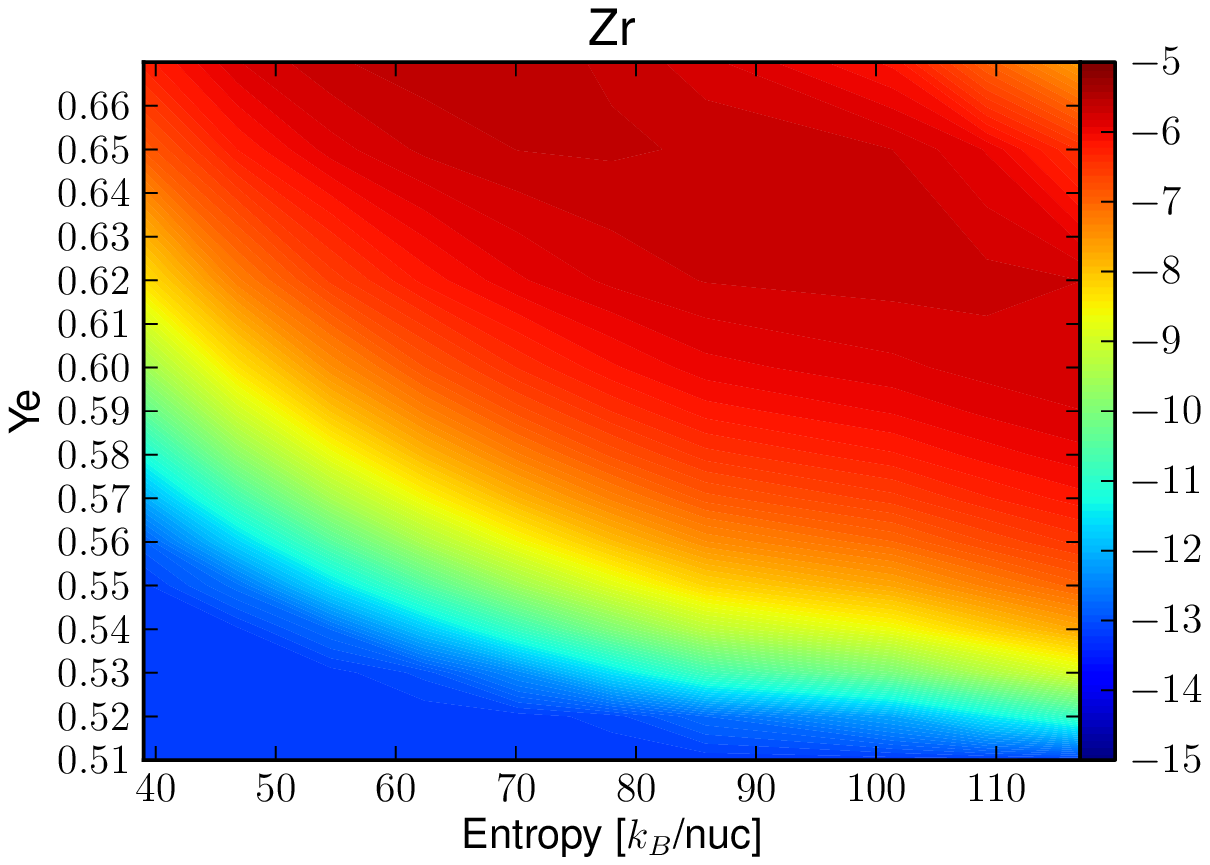}%
\includegraphics[width=0.5\linewidth]{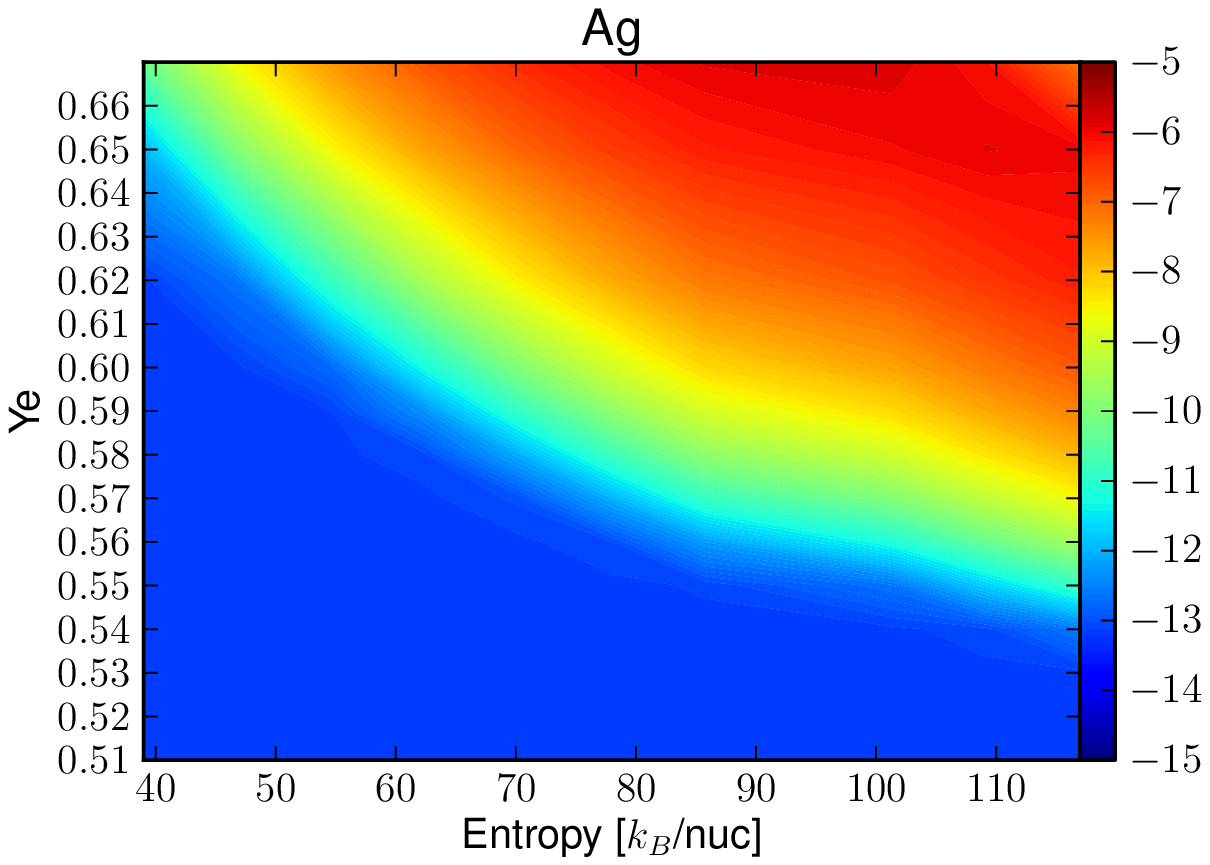}
\caption{Color contours show the abundance in log scale of Sr, Y, Zr,
  and Ag for different $Y_e$ and entropy.}
\label{fig:prich_elem_sye}
\end{center}
\end{figure*}

After studying neutron- and proton-rich wind nucleosynthesis, we find
two main differences in the final abundances. First, in proton-rich
conditions the abundance pattern is homogeneous. Depending on the wind
parameter the heaviest element that can be produced changes. However,
the overall pattern remains similar as it can be seen in
Fig.~\ref{fig:prich_elem_sye} where the abundance of Sr, Y, Zr, and Ag
are shown (color contours) for a broad range of entropies and electron
fractions. If we compare this figure to the left column in
Fig.~\ref{fig:nrich_elem_stau_ye}, the difference between neutron- and
proton-rich conditions becomes clear. In proton-rich winds the
abundances smoothly change when varying wind parameters, while in
neutron rich conditions this variations are far from been smooth and
linear.

 The second difference is the total abundance of heavy nuclei
 ejected. In proton-rich conditions the amount of heavy nuclei created
 is significantly smaller ($Y_{\mathrm{seed}}\lesssim 10^{-4}$) than
 in neutron-rich winds ($Y_{\mathrm{seed}}\gtrsim 10^{-3}$). The
 ejecta of proton-rich winds consist mainly of protons and alpha
 particles, while for neutron-rich the main contribution are alpha
 particles. This can be relevant for calculating the supernova
 contribution to the solar system abundances. For example it was shown
 by \cite{Hoffman.Woosley.etal:1996} that there is a limit on the
 amount of matter that can be ejected by every supernova with given
 electron fraction ($Y_e \lesssim 0.47$) in order not to over produce
 elements like Sr, Y, and Zr.

\section{Summary and conclusions}
\label{sec:summary}

The nucleosynthesis studies of neutrino-driven winds combined with
observations of chemical abundances can give important hints about
supernova explosions and the origin of elements beyond iron. Although
neutrino-driven winds have been extensively studied as the best
candidate for the production of heavy elements by the r-process,
recent simulations indicate that winds are in general not enough
neutron rich to produce elements heavier than silver. The production
and chemical evolution of elements between Sr and Ag remains an open
question and the contribution of neutrino-driven winds can reveal new
aspects about core-collapse supernova.

We have systematically studied the synthesis of lighter heavy elements
(between Sr and Ag) in slightly neutron-rich ($Y_e<0.5$) and
proton-rich ($Y_e>0.5$) winds. The nucleosynthesis in neutrino-driven
winds depends on entropy, expansion time scale, and electron
fraction. Therefore, we have varied all of these wind parameters and
analyzed their impact on the abundances of lighter heavy elements. The
entropy affects in a similar way for neutron- and proton-rich
conditions: higher entropies allow for more free nucleons and less
seed nuclei. This results in higher neutron-to-seed ratio that for
both conditions triggers the production of heavier elements. The
expansion time scale has the same effect as the entropy in
neutron-rich winds: faster expansion reduces the amount of seed
nuclei. However, for the $\nu p$-process a faster expansion not only
means less seed nuclei but also less time that the matter is exposed
to antineutrinos. The net result is less neutrons and less seed,
therefore variations of the overall expansion time scale do not lead
to large changes in  abundances. The impact of the electron
fraction has been discussed in previous works and we have summarized
that in neutron-rich conditions lower $Y_e$ and in proton-rich
conditions higher $Y_e$ lead to the formation of heavier elements.

In addition to the obvious difference in the isotopic composition,
there are two other important differences between neutron- and
proton-rich winds: the amount of matter ejected and the abundance
pattern. These characteristics can be key when we compare to
observations and study the contribution of supernova to the solar
system abundances.

In proton-rich winds, most of the ejected matter consists of protons
and alpha particles, while the amount of heavy nuclei is very
small. In neutron-rich conditions, alpha particles also dominate but
heavy nuclei are more abundant than in proton-rich conditions. Indeed
if every neutrino-driven wind stays neutron-rich for several seconds,
there would be an overproduction around $A=90$
\cite{Hoffman.Woosley.etal:1996, Witti.etal:1994}. This already
implies that not every neutrino-driven wind can be neutron-rich. Note
that for electron fractions very close to 0.5 the amount of heavy
elements ejected is again very small. Therefore, one could constrain
the neutron-richness assuming all winds are similar.

Other way of constraining the wind parameters is to compare the
abundance obtained in our nucleosynthesis studies to observations of
very old stars. There are several indications that the elements
between Sr and Ag are produced by a different site or process than the
heavier elements \cite{Qian.Wasserburg:2001, Montes.etal:2007}. The
fingerprint of this may be on stars with low enrichment of heavy
r-process elements \cite{Honda.etal:2006}. In these observations,
there are small variations in the abundance pattern for the lighter
heavy elements.  When we calculate the nucleosynthesis for different
wind parameters, we find that the pattern is rather robust for
proton-rich conditions but it varies significantly for neutron-rich
conditions. We have presented an overview of the dependency of the
abundance of Sr, Y, Zr, and Ag on the wind parameters. While in
proton-rich winds the abundance change smoothly when varying the wind
parameters, in neutron-rich winds we find strong variations in the
abundances for small changes in any of the wind parameters. Therefore,
if the neutrino-driven wind is responsible for the observed abundances
at low metallicities, neutron-rich conditions would not provide the
robustness observed in the abundance pattern. This may suggest that
the wind is not neutron or proton rich but a combination of both
conditions.

In order to make stronger conclusions for the wind parameters and its
neutron richness more nucleosynthesis studies, observations, and
nuclear experiments are necessary. All these will be achieved in the
near future. The nucleosynthesis processes (weak r-process and $\nu
p$-process) keep the path close to stability, therefore the relevant
experimental information can be obtained with future radioactive ion
beam facilities. To this end further nucleosynthesis studies are
required to explore the sensitivity of the still unmeasured nuclear
properties and thus to motivate the most relevant experiments. From
the observational perspective, further observations are necessary for
more stars with low enrichment of heavy r-process elements as well as
for more elements for the already observed low metallicity
stars. Combining all this with nucleosynthesis studies based on more
advanced supernova simulations and on parametric models, it will be
possible to constrain the neutron richness of the neutrino-driven
winds. Even if the conclusion were that no elements heavier than iron
are produced in core-collapse supernova, this will already strongly
limit the wind parameters and the neutrino energies and
luminosities. We are about to be able to use observations and
nucleosynthesis studies, like the one presented here, to learn more
about supernova explosions and neutrino-driven winds.

\ack 

This work was funded by the Helmholtz-University Young Investigator
grant No. VH-NG-825.  The authors acknowledge support to EuroGENESIS,
a collaborative research program of the European Science Foundation
(ESF).



\end{document}